
%
\input harvmac

\def\Titleh#1#2{\nopagenumbers\abstractfont\hsize=\hstitle\rightline{#1}%
\vskip .5in\centerline{\titlefont #2}\abstractfont\vskip .5in\pageno=0}

\def\CTPa{\it Center for Theoretical Physics, Department of Physics,
      Texas A\&M University}
\def\CTPb{\it College Station, TX 77843-4242, USA}
\def\HARCa{\it Astroparticle Physics Group,
Houston Advanced Research Center (HARC)}
\def\HARCb{\it The Woodlands, TX 77381, USA}

\def\ie{\hbox{\it i.e.}}     
\def\eg{\hbox{\it e.g.}}

\catcode`\@=11 

\def\lsim{\mathrel{\mathpalette\@versim<}}
\def\gsim{\mathrel{\mathpalette\@versim>}}
\def\@versim#1#2{\vcenter{\offinterlineskip
    \ialign{$\m@th#1\hfil##\hfil$\crcr#2\crcr\sim\crcr } }}
\def\boxit#1{\vbox{\hrule\hbox{\vrule\kern3pt
      \vbox{\kern3pt#1\kern3pt}\kern3pt\vrule}\hrule}}

\def\t1{{\tilde 1}}

\def\JL{J. L. Lopez}
\def\DVN{D. V. Nanopoulos}

\def\GeV{\,{\rm GeV}}

\def\ipb{\,{\rm pb^{-1}}}

\def\PLB#1#2#3{Phys. Lett. B {\bf#1} (19#2) #3}

\def\PRD#1#2#3{Phys. Rev. D {\bf#1} (19#2) #3}
\def\PRL#1#2#3{Phys. Rev. Lett. {\bf#1} (19#2) #3}
\def\PRT#1#2#3{Phys. Rep. {\bf#1} (19#2) #3}

\def\TAMU#1{Texas A \& M University preprint CTP-TAMU-#1}

\nref\LN{For a review see, A. B. Lahanas and \DVN, \PRT{145}{87}{1}.}
\nref\LNWZ{\JL, \DVN, X. Wang, and A. Zichichi, \TAMU{76/92}, CERN/LAA/92-023,
and CERN-PPE/92-194.}
\nref\LNPWZ{\JL, \DVN, H. Pois, X. Wang, and A. Zichichi, \TAMU{89/92},
CERN-TH.6773/93, and CERN/LAA/93-01.}
\nref\ANabc{R. Arnowitt and P. Nath, \PRL{69}{92}{725}; P. Nath and
R. Arnowitt, \PLB{287}{92}{89} and \PLB{289}{92}{368}.}
\nref\LNP{\JL, \DVN, and H. Pois, \TAMU{61/92} and CERN-TH.6628/92 (to appear
in Phys. Rev. D).}
\nref\LNPZ{\JL, \DVN, H. Pois, and A. Zichichi, \PLB{299}{93}{262}.}
\nref\LNZa{\JL, \DVN, and A. Zichichi, \PLB{291}{92}{255}.}
\nref\ANcosm{R. Arnowitt and P. Nath, \TAMU{65,66/92}, NUB-TH-3055,3066/92.}
\nref\LNZb{\JL, \DVN, and A. Zichichi, \TAMU{68/92}, CERN-TH.6667/92, and
CERN-PPE/92-188.}
\nref\DNh{M. Drees and M. M. Nojiri, \PRD{45}{92}{2482}.}
\nref\Hilgart{J. Hilgart, Talk presented at the 1993 Aspen Winter Conference.}
\nref\HHG{See \eg, {\it The Higgs Hunter's Guide}, J. Gunion, H. Haber, G.
Kane, and S. Dawson (Addisson-Wesley, Redwood City, 1990).}
\nref\ERZ{Y. Okada, M. Yamaguchi, and T. Yanagida, Prog. Theor. Phys.
{\bf85} (1991) 1 and \PLB{262}{91}{54}; J. Ellis, G. Ridolfi, and F. Zwirner,
\PLB{257}{91}{83}; H. Haber and R. Hempfling, \PRL{66}{91}{1815}.}
\nref\aspects{S. Kelley, \JL, \DVN, H. Pois, K. Yuan,  \TAMU{16/92} (to appear
in Nucl. Phys. B).}
\nref\KLNPYh{S. Kelley, \JL, \DVN, H. Pois, K. Yuan, \PLB{285}{92}{61}.}

\nfig\I{The cross section $\sigma(e^+e^-\to Z^*h\to\nu\bar\nu h$) as a function
of the lightest Higgs boson mass $m_h$ at the $Z$-pole for both supergravity
models. The arrows near 60 GeV indicate the improved lower bounds on $m_h$
(relative to the MSSM lower bound of 43 GeV), while the arrows near 70 GeV
indicate possible future lower bounds if LEPI were to establish a lower
bound on the SM Higgs boson mass of 70 GeV. The gaps on the curves are to be
understood as filled by intermediate points in the discrete parameter space
explored.}
\nfig\II{The ratio $f={\rm BR}(h\to2\,{\rm jets})_{\rm SUSY}/{\rm
BR}(h\to2\,{\rm jets})_{\rm SM}$ for both models as a function of $m_h$.}
\nfig\III{The contours of $\sin^2(\alpha-\beta)=0.99$ in generic supergravity
models in the $(m_{\tilde g},\tan\beta)$ plane for increasing values of
$\xi_0$. For $\xi_0>3$ all contours fall below the experimental lower bound
on $m_{\tilde g}$. The no-scale flipped model has $\xi_0=0$, while the
minimal model requires $\xi_0\gsim3$. Since areas to the right of the contours
have yet larger values of $\sin^2(\alpha-\beta)$, the plot shows the approach
to the SM-like limit of the Higgs sector (\ie, $\sin^2(\alpha-\beta)\to1$) for
increasingly larger gluino masses. This trend is accelerated by increasing
$\xi_0$ values. (For the $\xi_0=0$ case consistency conditions impose an upper
bound on $\tan\beta\,(\lsim15)$ \aspects.)}

\centerline{EUROPEAN ORGANIZATION FOR NUCLEAR RESEARCH}
\bigskip
\Titleh{\vbox{\baselineskip12pt
\hbox{CERN-PPE/93--17}
\hbox{8 February, 1993}
\hbox{CERN/LAA/93--04}
\hbox{CTP--TAMU--05/93}\hbox{ACT--1/93}}}
{\vbox{\centerline{Improved LEP Lower Bound on the Lightest SUSY}
\centerline{Higgs Mass from Radiative Electroweak Breaking}
\centerline{and its Experimental Consequences}}}
\centerline{JORGE~L.~LOPEZ$^{(a)(b)}$, D.~V.~NANOPOULOS$^{(a)(b)}$,
H. POIS$^{(a)(b)}$,}
\centerline{XU WANG$^{(a)(b)}$, and A. ZICHICHI$^{(c)}$}
\smallskip
\centerline{$^{(a)}$\CTPa}
\centerline{\CTPb}
\centerline{$^{(b)}$\HARCa}
\centerline{\HARCb}
\centerline{$^{(c)}${\it CERN, Geneva, Switzerland}}
\vskip .2in
\centerline{ABSTRACT}
We show that the present LEPI lower bound on the Standard Model Higgs boson
mass ($M_H\gsim60\GeV$) applies as well to the lightest Higgs boson ($h$) of
the minimal $SU(5)$ and no-scale flipped $SU(5)$ supergravity models. This
result would persist even for the ultimate LEPI lower bound ($M_H\gsim70\GeV$).
We show that this situation is a consequence of a
decoupling phenomenon in the Higgs sector driven by radiative electroweak
breaking for increasingly larger sparticle masses, and thus it should
be common to a large class of supergravity models.
 A consequence of $m_h\gsim60\GeV$ in the minimal $SU(5)$ supergravity model is
the exclusion from the allowed parameter space of `spoiler modes'
($\chi^0_2\to\chi^0_1 h$) which would make the otherwise very promising
trilepton signal in $p\bar p\to\chi^\pm_1\chi^0_2X$ unobservable at Fermilab.
Within this model we also obtain stronger upper
bounds on the lighter neutralino and chargino masses, \ie,
$m_{\chi^0_1}\lsim50\GeV$, $m_{\chi^0_2,\chi^\pm_1}\lsim100\GeV$.
This should encourage experimental searches with existing facilities.
\bigskip
\Date{}

\newsec{Introduction}
The current renewed interest on supersymmetry and its phenomenological
consequences can, within the context of the MSSM, only go so far. This
limitation is due to the large size (at least 21-dimensional) of the
parameter space that should be explored. In practice people routinely impose
certain `grand unification-' and `supergravity-inspired' relations among the
model parameters, although usually not in a completely consistent way and
omitting several equally well motivated constraints. This hodgepodge approach
to minimizing the number of assumptions in order to get the most
`model-independent' results can be misleading.\foot{A relevant
example of this occurs in the Higgs sector where the decay mode $h\to AA$ can
be a spectacular signature in the MSSM but has been shown to be forbidden in
supergravity models \DNh.} Supergravity models with radiative breaking of the
electroweak symmetry \LN\ have a much reduced parameter space (three
supersymmetry breaking parameters ($m_{1/2},m_0,A$), $\tan\beta$, and the
top-quark mass) and are therefore highly predictive and falsifiable. We have
recently studied the experimental signatures \refs{\LNWZ,\LNPWZ} for two such
models: (i) the minimal $SU(5)$ supergravity model including the stringent
constraints of proton stability \refs{\ANabc,\LNP,\LNPZ} and cosmology
\refs{\LNZa,\LNP,\LNPZ,\ANcosm}, and (ii) the no-scale flipped $SU(5)$
supergravity model \LNZb. In this note we focus on the constraints from current
LEPI data on the lightest Higgs boson mass in these two models. We show that
due to the nature of the Higgs masses, couplings, and branching fractions, at
LEPI for $m_h\lsim70\GeV$ the $h$ particle should be basically
indistinguishable from the Standard Model Higgs boson, and
therefore the experimental lower bound on the latter should also apply  to the
former. We then explore the generality of this result and give arguments, based
on the  built-in radiative electroweak breaking mechanism, for its validity
in a more general class of supergravity models. The pre-existing severe limits
on the minimal $SU(5)$ supergravity model (based on $m_h>43\GeV$) are shown to
be even stronger, implying that LEPII and Fermilab should be able to explore a
yet larger portion (if not all) of the parameter space of this model.
\newsec{Improved lower bounds on $m_h$}
The current LEPI lower bound on the Standard Model (SM) Higgs boson mass
($m_H>61.6\GeV$ \Hilgart) is obtained by studying the
process $e^+e^-\to Z^*H$ with subsequent Higgs decay into two jets. The MSSM
analog of this production process leads to a cross section differing just by a
factor of $\sin^2(\alpha-\beta)$, where $\alpha$ is the SUSY Higgs mixing angle
and $\tan\beta=v_2/v_1$ is the ratio of the Higgs vacuum expectation values
\HHG. The published LEPI lower bound on the lightest SUSY Higgs boson mass
($m_h>43\GeV$) is the result of allowing $\sin^2(\alpha-\beta)$ to vary
throughout the MSSM parameter space and by considering the $e^+e^-\to Z^*h,hA$
cross sections. It is therefore possible that in specific models (which embed
the MSSM), where $\sin^2(\alpha-\beta)$ is naturally restricted to be near
unity (as for example discussed in the next section), the lower bound on $m_h$
could rise, and even reach the SM lower bound if ${\rm BR}(h\to2\,{\rm jets})$
is SM-like as well. This we will show is the case for the
two supergravity models in hand.

Non-observation of a SM Higgs signal puts the following upper bound in the
number of expected 2-jet events.
\eqn\I{\#{\rm events}_{\rm\,SM}=\sigma(e^+e^-\to Z^*H)_{\rm SM}\times
{\rm BR}(H\to2\,{\rm jets})_{\rm SM}\times\int{\cal L}dt<3.}
The SM value for ${\rm BR}(H\to2\,{\rm jets})_{\rm SM}\approx
{\rm BR}(H\to b\bar b+c\bar c+gg)_{\rm SM}\approx0.92$ \HHG\ corresponds to
an upper bound on $\sigma(e^+e^-\to Z^*H)_{\rm SM}$. Since this is a
monotonically decreasing function of $m_H$, a lower bound on $m_H$ follows,
\ie, $m_H>61.6\GeV$ as noted above. We denote by $\sigma_{\rm SM}(61.6)$ the
corresponding value for $\sigma(e^+e^-\to Z^*H)_{\rm SM}$. For the MSSM the
following relations hold
\eqna\A
$$\eqalignno{\sigma(e^+e^-\to Z^*h)_{\rm
SUSY}&=\sin^2(\alpha-\beta)\sigma(e^+e^-\to
Z^*H)_{\rm SM},&\A a\cr
{\rm BR}(h\to2\,{\rm jets})_{\rm SUSY}&=f\cdot{\rm BR}(H\to2\,{\rm jets})_{\rm
SM}.&\A b\cr}$$
{}From Eq. \I\ one can deduce the integrated luminosity achieved,\break
$\int{\cal L}dt=3/(\sigma_{\rm SM}(61.6){\rm BR}_{\rm SM})$. In analogy with
Eq. \I, we can write
\eqn\II{\#{\rm events}_{\rm\,SUSY}=\sigma_{\rm SUSY}(m_h)\times{\rm BR}_{\rm
SUSY}\times\int{\cal L}dt=3f\cdot\sigma_{\rm SUSY}(m_h)/\sigma_{\rm
SM}(61.6)<3.}
This immediately implies the following condition for {\it allowed} points in
parameter space
\eqn\III{f\cdot\sigma_{\rm SUSY}(m_h)<\sigma_{\rm SM}(61.6).}
The cross section $\sigma_{\rm SUSY}(m_h)$ is shown in Fig. 1 for both models.
The values shown for the minimal $SU(5)$ model also correspond to the SM
result since one can verify that $\sin^2(\alpha-\beta)>0.9999$ in this case.
For the flipped model there is a hard-to-see ($\sin^2(\alpha-\beta)>0.95$)
drop relative to the SM result (as shown on the top row plots) for some points.
The ratio $f$ versus $m_h$ is shown
in Fig. 2.\foot{In the calculation of ${\rm BR}(h\to2\,{\rm jets})_{\rm SUSY}$
we have included {\it all} contributing modes, in particular the invisible
$h\to\chi^0_1\chi^0_1$ decays.} It is interesting to remark that the two models
differ little from the SM and in fact the proper lower bound on $m_h$
(which follows from the use of Eq. \III) is marked by the set of arrows near 60
GeV in Fig. 1. Note that the bound is lowest (given by the left-most arrow
of the four near 60 GeV in Fig. 1) for the minimal $SU(5)$ ($\mu<0$)
since the $f$ ratio is smallest in this case. A similar analysis shows that if
$M_H>70\GeV$ is established at LEPI, then $m_h\gsim70\GeV$ would also follow
(note the second set of arrows in Fig. 1 around 70 GeV). The present bound
on $M_H$ has been obtained with $\approx3\times10^6$ hadronic $Z$-decays, which
is nearing the ultimate number achievable at LEPI. Therefore if Higgs events
are at all seen, LEPI should not be able to distinguish these two models from
each other or from the SM. Such a differentiation would require a detailed
study of the branching fractions \LNPWZ.
\newsec{Radiative breaking and decoupling}
The results in the previous section may signal a general feature of
supergravity models. In fact, the two examples considered above can be taken as
extreme cases of generic supergravity models with radiative electroweak
breaking, in that the no-scale flipped model has $\xi_0=m_0/m_{1/2}=0$, whereas
the minimal model has $\xi_0\gsim3$. The Higgs sector of the MSSM is generally
specified at tree-level by the arbitrary choice of two parameters,
\eg, $\tan\beta$ and $m_A$. At one-loop the whole spectrum enters, but for the
present purposes it will suffice to describe it in terms $m_t$ and $m_{\tilde
q}$ \ERZ. It is well known that in the limit
of $m_A\gg M_Z$ one recovers a SM-like theory for the $h$ the Higgs couplings
to fermions and vector bosons, with $m_h$ at its maximum
value ($m^2_h\approx \cos^22\beta M^2_Z+\Delta m^2_h$) and
$\sin^2(\alpha-\beta),|\sin\alpha/\cos\beta|,|\cos\alpha/\sin\beta|\approx1$.
Also, the $H,A,H^+$ decouple: they become increasingly heavy, degenerate, and
their couplings to fermions and vector bosons are driven to zero.
Since in the MSSM $m_A$ is a physical input parameter, there is no a priori
preferred value; experimentally, $m_A>23\GeV$. What mechanism may enforce
$m_A\gg M_Z$ is a question beyond the MSSM. On the other hand, radiative
electroweak breaking in supergravity models determines $m^2_A$ in terms of
other sectors  of the theory,
\eqn\IV{m^2_A=-{2\mu B\over\sin2\beta}+\Delta m^2_A
=m^2_{H_1}+m^2_{H_2}+2\mu^2+\Delta m^2_A,}
where $\mu,B,m^2_{H_1},m^2_{H_2}$ are parameters in the Higgs potential
(see \eg, Ref. \aspects), and $\Delta m^2_A$ represents the one-loop
correction. Since the renormalization group equations (RGEs) [which determine
$m_{H_1},m_{H_2}$ and $\mu$] scale with $m_{1/2}\propto m_{\tilde g}$ \aspects,
increasing $m_{\tilde g}$ will drive $m_A$ to larger values and the Higgs
sector to the SM-like limit. Furthermore, if the initial conditions for the
$m^2_{H_{1,2}}$ RGEs at the unification scale, \ie, $m_0=\xi_0 m_{1/2}$ are
increased, the stated behavior should be accelerated: decoupling should be
approached for lower values of $m_{\tilde g}$. To verify these qualitative
statements we have studied a class of minimal supergravity models with
$\xi_0=0,1,2$ and determined the
$\sin^2(\alpha-\beta)=0.99$ contours in the $(m_{\tilde g},\tan\beta)$
plane.\foot{The results are not sensitive to the choice of $\xi_A, m_t$, or the
sign of $\mu$, except for the portion of the contours which may become
phenomenologically excluded.} These contours are shown in Fig. 3 (points to
the right (left) of a given contour have larger (smaller) values of
$\sin^2(\alpha-\beta)$) and help to quantify our previous qualitative remarks,
and to explain the behavior observed in the two sample models considered in
Sec. 2. In Fig. 3 contours for $2<\xi_0<2.8$ occur to the left of the shown
$\xi_0=2$ contour, but still have some points with theoretically and
experimentally allowed values of the gluino mass. For $\xi_0>3$ contours
have no points for allowed values of $m_{\tilde g}$. Therefore, {\it any}
minimal supergravity model with $\xi_0>3$, implies $\sin^2(\alpha-\beta)>0.99$.
Precise statements about lower bounds on $m_h$ depend on ${\rm BR}(h\to2\,{\rm
jets})$ which is quite model dependent (see for example Fig. 2). However, the
only deviation from SM rates will arise from loop-induced decay modes (\eg,
$h\to gg$) and non-SM final states (\eg, $h\to\chi^0_1\chi^0_1$).

The point to be stressed is that if the supersymmetric Higgs sector is
found to be SM-like, this could be taken as indirect evidence for an
underlying radiative electroweak breaking mechanism,\foot{Or as a `cosmic
conspiracy', whichever one likes better.} since no insight could be garnered
from the MSSM itself.

\newsec{Experimental consequences for the minimal $SU(5)$ supergravity model}
The improved bound $m_h\gsim60\GeV$ mostly restricts low values of $\tan\beta$
and therefore the minimal $SU(5)$ supergravity model where $\tan\beta\lsim3.5$
\LNPZ. (The no-scale flipped $SU(5)$ supergravity model is also constrained for
small $\tan\beta$, but since in this model $\tan\beta$ can be as large as 32,
only a small region of parameter space is affected.)
In Ref. \LNPZ\ we obtained upper bounds on the light particle masses in the
minimal $SU(5)$
supergravity model ($\tilde g,h,\chi^0_{1,2},\chi^\pm_1$) for $m_h>43\GeV$. In
particular, it was found that $m_{\chi^\pm_1}\gsim100\GeV$ was only possible
for $m_h\lsim50\GeV$. The improved bound on $m_h$ immediately implies the
following considerably stronger upper bounds

\eqna\V
$$\eqalignno{m_{\chi^0_1}&\lsim52(50)\GeV, &\V a\cr
m_{\chi^0_2}&\lsim103(94)\GeV,&\V b\cr
m_{\chi^\pm_1}&\lsim104(92)\GeV,&\V c\cr
m_{\tilde g}&\lsim320\,(405)\GeV,&\V d\cr}$$
for $\mu>0$ ($\mu<0$). Imposing $m_h\gsim70\GeV$ does not change these results.

A related consequence is that the mass relation $m_{\chi^0_2}>m_{\chi^0_1}+m_h$
is not satisfied for any of the remaining points in parameter space
and therefore the $\chi^0_2\to\chi^0_1 h$ decay mode is not kinematically
allowed. Points where such mode was previously allowed (see the `+' symbols in
Fig. 2 in Ref. \LNPZ) led to a vanishing
trilepton signal in the reaction $p\bar p\to\chi^\pm_1\chi^0_2$ at Fermilab
(thus the name `spoiler mode') \LNWZ. The improved situation now implies
at least one event per $100\ipb$ for all remaining points in parameter space.
\newsec{Conclusions}
We have shown that the current experimental lower bound on the SM Higgs boson
mass can be used to impose constraints on the Higgs sector of supergravity
models, which are stronger than those possible in the generic MSSM. This is
the direct result of the underlying radiative electroweak symmetry breaking
mechanism which links the Higgs sector to the sparticle sector of the theory.
In fact, such a link leads to the SM-like limit of the Higgs sector in a
natural way for $\xi_0>3$ in any minimal supergravity model, since independent
gluino searches have obtained a lower limit to the scale involved in radiative
electroweak breaking. In the no-scale case ($\xi_0=0$), the possibility for
significant deviations from the SM-like limit exists \KLNPYh\ but requires
$m_h\gsim80\GeV$ and large $\tan\beta$. This region is obviously beyond the
reach of LEPI, but is accessible at LEPII \LNPWZ. We have pursued the
consequences of these ideas explicitly in two realistic supergravity models,
and have obtained rather stringent indirect constraints on the lighter
neutralino and chargino masses of the minimal $SU(5)$ supergravity model with
the new $h$ mass limit $m_h\gsim60\GeV$. We conclude that well motivated
theoretical assumptions open the way to observe experimentally a SM-like Higgs
boson.
\bigskip
\bigskip
\bigskip
\noindent{\it Acknowledgments}: J. L. would like to thank S. Katsanevas and
J. Hilgart for very helpful discussions. This work has been supported in part
by DOE grant DE-FG05-91-ER-40633. The work of J.L. has been supported by an SSC
Fellowship. The work of  D.V.N. has been supported in part by a grant from
Conoco Inc. The work of X. W. has been supported by a T-1 World-Laboratory
Scholarship. We would like to thank the HARC Supercomputer Center for the use
of their NEC SX-3 supercomputer and the Texas A\&M Supercomputer Center for the
use of their CRAY-YMP supercomputer.
\footatend\bigskip\bigskip\bigskip\immediate\closeout\rfile\writestoppt
\baselineskip=14pt\centerline{{\bf References}}\bigskip{\frenchspacing%
\parindent=20pt\escapechar=` \input refs.tmp\vfill\eject}\nonfrenchspacing

\listfigs
\bye